# Ptychographic modulation engine (PME): a low-cost DIY microscope add-on for coherent super-resolution imaging

Zichao Bian[1,3], Shaowei Jiang[1,3], Pengming Song[2], He Zhang[1], Pouria Hoveida[1], Kazunori Hoshino[1], and Guoan Zheng[1,2]

[1] Department of Biomedical Engineering, University of Connecticut, Storrs, CT, 06269, USA
[2] Department of Electrical and Computer Engineering, University of Connecticut, CT, 06269, USA
[3] These authors contributed equally to this work

E-mail: S.J. (shaowei.jiang@uconn.edu)

## Abstract

Imaging of biological cells and tissues often relies on fluorescent labels, which offer high contrast with molecular specificity. The use of exogenous labeling agents, however, may alter the normal physiology of the bio-specimens. Complementary to the established fluorescence microscopy, label-free quantitative phase imaging provides an objective morphological measurement tool for bio-specimens and is free of variability introduced by contrast agents. Here we report a simple and low-cost microscope add-on, termed Ptychographic Modulation Engine (PME), for super-resolution quantitative phase imaging. In this microscope add-on module, we attach a diffuser to a 3D-printed holder that can be mechanically moved to different x-y positions. We then use two vibrational motors to introduce random positional shifts to the diffuser. The add-on module can be placed between the objective lens and the specimen in most existing microscope platforms. Thanks to the diffuser modulation process, the otherwise inaccessible high-resolution object information can now be encoded into the captured images. In the ptychographic phase retrieval process, we jointly recover the complex object wavefront, the complex diffuser profile, and the unknown positional shifts of the diffuser. We demonstrate a 4-fold resolution gain over the diffraction limit of the employed 2X objective lens. We also test our approach for in-vivo cell imaging, where we are able to adjust the focus after the data has been captured. The reported microscope add-on provides a turnkey solution for super-resolution quantitative phase imaging. It may find applications in label-free bio-imaging where both large field-of-view and high resolution are needed.

## 1. Introduction

Imaging of biological cells and tissues often relies on fluorescent labels, which offer high contrast with molecular specificity. The use of exogenous labeling agents, however, may alter the normal physiology of the bio-specimens. As a result, label-free imaging techniques have attracted much attention from researchers in recent years [1-23]. One important class of label-free imaging techniques is quantitative phase imaging (QPI), which generates a quantitative map of the optical-path delay through the specimen. QPI provides an objective morphological measurement of the sample and is free of variability introduced by contrast agents. QPI can be implemented via various platforms and devices, from lensless instruments to microscope systems, with different methods of extracting the quantitative phase information [4-23]. Its applications in biomedicine can be found in a recent review paper [22].

In this work, we focus on a specific QPI approach termed ptychography. It is a lensless imaging approach that was originally proposed for electron microscopy and brought to fruition by Faulkner and Roderburg [24]. The goal of ptychography is to recover the phase information from diffraction intensity measurements at the far-field. In a typical implementation, a ptychographic dataset is obtained by collecting a sequence of diffraction patterns in which the object is mechanically scanned through a spatially confined illumination spot, termed 'probe'. This confined illumination spot limits the physical extent of the object for each diffraction pattern measurement and serves as a support constraint for the recovery process. The acquired 2D images are then processed into an estimate of the sample's complex transmission profile.

The reconstruction process of ptychography has its origins in the phase retrieval method first proposed by Gerchberg and Saxton [25]. It iteratively seeks a solution that is consistent with the measured diffraction patterns. In the spatial domain, the finite extent of the illumination spot serves as the support constraint for the solution. In the Fourier domain, the diffraction intensity measurement can be used to update the amplitude component of the solution, while the phase component is kept unchanged. The phase retrieval process is then repeated for all diffraction measurements and repeated until the solution converges.

Ptychography's unique procedure has led to the generation of many impressive X-ray and electron microscope instruments that defy the conventional resolution limits of their detectors and focusing elements [26-32]. In particular, ptychography has become an indispensable imaging tool in most synchrotron X-ray facilities worldwide.

A typical ptychographic implementation does not use any lens. We have developed a lens-based ptychography approach termed Fourier ptychography [19, 33-35]. In Fourier ptychography, we use an LED array to sequentially illuminate the sample from multiple incident angles. At each illumination angle, it records a low-resolution intensity image through the low numerical aperture (NA) objective lens. The objective's pupil function, therefore, imposes a circular constraint in the Fourier domain. Scanning the incident angle of the illumination beam effectively scans the aperture constraint in the Fourier domain. By employing the same phase retrieval process as ptychography, we can synthesize the aperture support in the Fourier domain to improve the achievable resolution in the spatial domain, enabling a combination of high resolving power and large field of view at the same time

[19, 33, 34, 36-39]. Moreover, both quantitative absorption and phase contrast can be reconstructed simultaneously. Analogous to the probe retrieval in regular ptychography, Fourier ptychography can also recover the optical transfer function, i.e., the spatial-frequency response of the objective, which allows aberrations to be reliably corrected for [34, 40].

One major limitation of Fourier ptychography is the thin sample requirement for angle-varied illumination. Only under this requirement will the low-resolution images obtained at different incident angles uniquely map to different passbands of the 2D Fourier spectrum, allowing the phase retrieval algorithm to impose the panning spectrum constraint to recover a high-resolution complex sample image. If the thin sample requirement is not met, this one-to-one mapping relationship in the Fourier plane is invalid, and the panning spectrum constraint cannot be imposed [41-43].

To circumvent the thin sample requirement noted above, we have demonstrated an approach termed aperture-scanning Fourier ptychography, where the sample is illuminated with a single plane wave and an aperture is mechanically scanned to different positions at the Fourier plane [41]. A key innovation of aperture-scanning Fourier ptychography is to perform lightwave modulation at the detection path. As such, the recovered image depends upon how the complex wavefront exits the sample – not enters it. Therefore, the sample thickness becomes irrelevant during reconstruction. After recovery, the exiting complex wavefront can be back-propagated to any plane along the optical axis for 3D refocusing. This approach, however, cannot improve the resolution beyond the diffraction limit imposed by the objective lens. Therefore, it is not a microscopy approach to achieve both large field of view and high resolution as that of the original Fourier ptychography.

By integrating the detection-path-modulation concept [41] with the innovation of scattering lens [44, 45], we have recently developed a super-resolution imaging approach termed ptychographic structured modulation [46]. In this approach, we place a thin diffuser (i.e., a scattering lens) in between the sample and the objective lens to modulate the complex light waves from the object. The otherwise inaccessible high-resolution object information can thus be encoded into the captured images thanks to the diffuser modulation.

A limitation of the ptychographic structured modulation approach is the need for precise mechanical scanning of the diffuser. Here we report the development of a simple and low-cost microscope add-on module, termed ptychographic modulation engine (PME), to convert the ptychographic structured modulation concept into a turnkey solution for most existing microscope platforms. In particular, we have developed a ptychographic phase retrieval procedure to recover the super-resolution object wavefront, the diffuser profile, and the unknown positional shift of the diffuser.

Therefore, precise mechanical scanning that is critical in conventional ptychography experiments is no longer needed in our implementation. In the reported microscope add-on module, we attach a diffuser to a 3D-printed holder that can be mechanically moved to different x-y positions. Two low-cost vibration motors are then used to introduce random positional shifts to the diffuser.

In the following, we will first discuss the design considerations for the PEM add-on module. We will then discuss and validate our phase retrieval procedures where the positional shifts are treated as unknown parameters. Next, we will demonstrate the imaging performance of the reported add-on via resolution targets, fixed microscope slides, and in-vivo cell imaging experiment. We demonstrate a 4-fold resolution gain over the diffraction limit of the employed 2X objective lens. For in-vivo cell imaging, we also show that the focus can be computationally adjusted after the data has been captured. The reported microscope add-on provides a turnkey solution for super-resolution quantitative phase imaging. It may find applications in label-free bio-imaging where both large field-of-view and high resolution are needed.

## 2. Ptychographic modulation engine (PME)

Figure 1 shows the optical scheme and our design of the PME add-on module, which can be placed between the objective lens and the sample. As such, the light waves from the object will be modulated by the diffuser and the high-resolution object information otherwise inaccessible is now encoded in the captured images (Fig. 1(a)). The final achievable resolution is not limited by the NA of the objective lens. Instead, it is limited by the feature size of the thin diffuser. In our experiment, we demonstrate a 4-fold resolution gain beyond the diffraction limit of the employed objective lens.

Figure 1(b) shows the 3D-printed design of our PME module. It comprises two sets of parallel-beam flexures to generate motion in the x-y plane. Each set of flexure is designed to be compliant in one direction (either x- or y-axis direction) but is stiff in the other motion directions. This design allows relatively easy x-y translational motion while constraining the unwanted axial or rotational motion. An eccentric rotating mass vibrational motor is used for each set of the flexure to actuate the stage in its corresponding direction, as shown in Fig. 1(b). Other liner resonant actuators can also be used for this purpose.

The entire PME add-on module can be placed on a regular microscope platform for super-resolution imaging. Figure 2(a) shows our experimental setup where the PME add-on is placed on an upright microscope with a 2X, 0.055 NA objective lens. The diffuser is made by coating ~1-μm microspheres on a coverslip. In our PME design, we choose the width and the length of the parallel-beam flexures to ensure that the maximum displacement is around ~50 μm in both x and y directions. A larger maximum displacement leads to a larger unknown diffuser profile for recovery. A smaller displacement, on the other hand, may not introduce enough x-y positional shift for the ptychographic modulation process. The height of the flexures is ~1 cm to avoid possible motion along the axial direction.

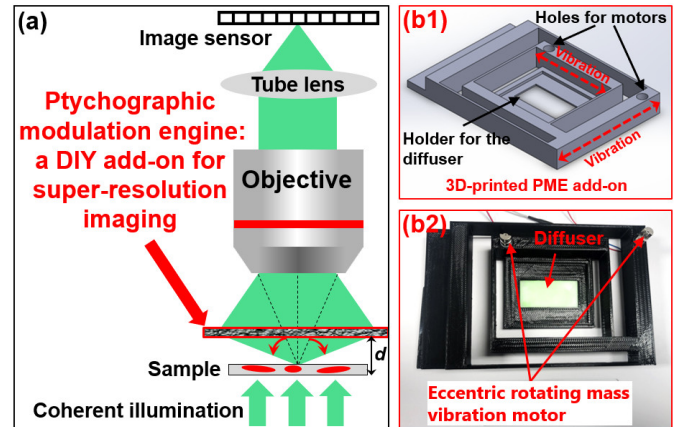

**Figure 1**. Ptychographic modulation engine (PME): a low-cost DIY microscope add-on for super-resolution imaging. (a) The optical scheme of the reported microscope add-on. (b1) The 3D design of the PME add-on, where the holder can be mechanically moved to different x-y positions via motor vibration. (b2) The 3D-printed PME prototype, where we use two eccentric rotating mass vibration motors to introduce random positional shift for the diffuser.

To operate the PME add-on module, we can simply apply a voltage to the vibrational motors and acquire the diffuser modulated images through the objective lens (supplementary video 1). The positional shifts of the diffuser are treated as unknown parameters in this process. We can then recover the super-resolution object wavefront, the complex diffuser profile, the positional shifts of the diffuser. Figure 2(b) shows the recovered positions of the diffuser using our phase retrieval procedure.

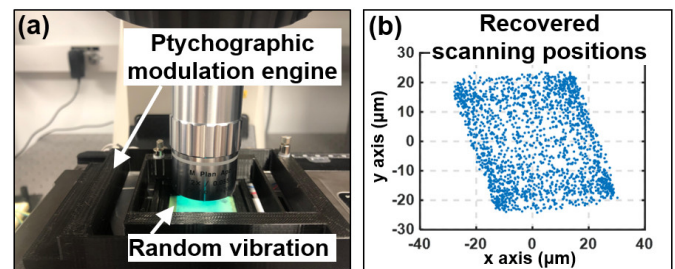

**Figure 2**. (a) The PME add-on module can be attached to an existing microscope platform for super-resolution microscopy imaging (also refer to supplementary video 1). (b) The recovered scanning positions of the diffuser using the reported phase retrieval procedure discussed in Section 3.

## 3. Recovery procedures

The forward imaging model of the reported system can be written as
$$I_j(x,y) = \left|W(x,y) * PSF_{free}(d) \cdot D(x - x_j, y - y_j) * PSF_{free}(-d) * PSF_{obj}\right|^2, \quad (1)$$
where $I_j(x,y)$ is the $j^{th}$ intensity measurement ($j = 1,2,...,J$), $W(x,y)$ is the complex exit wavefront of the object, $D(x,y)$ is the complex profile of the diffuser, $(x_j, y_j)$ is the $j^{th}$ positional shift of the diffuser, '·' stands for point-wise multiplication, and '*' denotes convolution operation. In Eq. (1), '$d$' is the distance between the exit wavefront and the diffuser. We use $PSF_{free}(d)$ to model the point spread function (PSF) for free-space propagation over distance '$d$'. Similarly, we use $PSF_{obj}$ to model the low-pass filtering process of the objective lens and its Fourier transform represents the coherent transfer function (CTF) of the microscope system.

We first initialize the object wavefront $W(x,y)$, the complex diffuser profile $D(x,y)$, and the unknown positional shifts $(x_j, y_j)$s as follows:
$$W(x,y) = \frac{1}{J}\sum_{j=1}^{J}\sqrt{I_j(x,y)}, \quad (2)$$

$$(x_j, y_j) = arg \max_{(x_j,y_j)} \frac{\sqrt{I_1(x,y)}}{\sum_{j=1}^{J}\sqrt{I_j(x,y)}} \star \frac{\sqrt{I_j(x,y)}}{\sum_{j=1}^{J}\sqrt{I_j(x,y)}}, \quad (3)$$

$$D(x,y) = \frac{1}{J}\sum_{j=1}^{J}\sqrt{I_j(x+x_j, y+y_j)}. \quad (4)$$

In Eq. (3), '$\star$' denotes the cross-correlation operation.

Our reconstruction process is based on the ptychographic phase retrieval process. The detailed procedures of the reconstruction algorithm designed specifically for the reported PME add-on module are as follows:

1. Obtain the low-resolution diffuser profile, $D_{low}(x,y)$, via
$$D_{low}(x,y) = |D(x,y) * PSF_{free}(-d) * PSF_{obj}|. \quad (5)$$
We can then update the positional shift by maximizing the following image cross-correlation:
$$(x_j, y_j) = arg \max_{(x_j,y_j)} D_{low}(x,y) \star \sqrt{I_j(x,y)} \quad (6)$$

2. Propagate the object wavefront to the diffuser plane:
$$W^{prop}(x,y) = W(x,y) * PSF_{free}(d), \quad (7)$$
where $W^{prop}(x,y)$ is the object wavefront after free-space propagation of distance '$d$'.

3. The shifted diffuser profile is given as
$$D_j(x,y) = D(x - x_j, y - y_j). \quad (8)$$

4. An estimate of the exit wave leaving the diffuser is given by the product of the shifted diffuser profile and the object wavefront after propagation:
$$\phi_j(x,y) = W^{prop}(x,y) \cdot D_j(x,y). \quad (9)$$

5. The exit wave $\phi_j(x,y)$ is low-pass filtered by the defocused CTF in the Fourier domain to obtain $\psi_j(x,y)$. The defocused CTF is calculated based on the distance '$d$' between the exit wavefront and the diffuser.
$$\begin{aligned}\psi_j(x,y) &= \mathcal{F}^{-1}\{\Psi(k_x, k_y)\} \\ &= \mathcal{F}^{-1}\{\Phi_j(k_x, k_y) \cdot CTF_{defocus}\} \\ &= \mathcal{F}^{-1}\{\mathcal{F}(\phi_j(x,y)) \cdot CTF_{defocus}\}.\end{aligned} \quad (10)$$

6. Apply the modulus constraint to obtain the updated $\Psi'(k_x, k_y)$,
$$\begin{aligned}\Psi'(k_x, k_y) &= \mathcal{F}\{\psi_j'(x,y)\} \\ &= \mathcal{F}\left\{\frac{\psi_j(x,y)}{|\psi_j(x,y)|} \cdot \sqrt{I_j(x,y)}\right\}.\end{aligned} \quad (11)$$

7. Update the Fourier spectrum of the exit wave $\phi_j(x,y)$ in the Fourier domain using the ePIE algorithm [47]:
$$\Phi_j'(k_x, k_y) = \Phi_j(k_x, k_y) + \beta_\Phi \frac{conj(CTF_{defocus})\{\Psi'(k_x, k_y) - \Psi(k_x, k_y)\}}{|CTF_{defocus}|^2_{max}}, \quad (12)$$
where conj(·) represents complex conjugate and $\beta_\Phi$ is the step size.

8. The updated exit wave can be expressed as
$$\phi_j'(x,y) = \mathcal{F}^{-1}\{\Phi_j'(k_x, k_y)\}. \quad (13)$$

9. Based on the updated exit wave $\phi_j'(x,y)$, we then update the object wavefront and the shifted diffuser profile using the newly developed rPIE algorithm [48]:
$$W_{update}^{prop}(x,y) = W^{prop}(x,y) + \frac{conj(D_j(x,y)) \cdot \{\phi_j'(x,y) - \phi_j(x,y)\}}{(1 - \alpha_{obj})|D_j(x,y)|^2 + \alpha_{obj}|D_j(x,y)|^2_{max}}, \quad (14)$$

$$D_{j,update}(x,y) = D_j(x,y) + \frac{conj(W_{update}^{prop}(x,y)) \cdot \{\phi_j'(x,y) - \phi_j(x,y)\}}{(1 - \alpha_D)|W_{update}^{prop}(x,y)|^2 + \alpha_D|W_{update}^{prop}(x,y)|^2_{max}}. \quad (15)$$
Here $\alpha_{obj}$ and $\alpha_D$ are two constants that can tune the convergence of the algorithm.

10. Update the object wavefront by propagating $W_{update}^{prop}(x,y)$ for distance '$-d$':
$$W_{update}(x,y) = W_{update}^{prop}(x,y) * PSF_{free}(-d). \quad (16)$$

11. The diffuser profile can be updated by shifting the $D_{j,update}(x,y)$ by $(-x_j, -y_j)$:
$$D_{update}(x,y) = D_{j,update}(x + x_j, y + y_j). \quad (17)$$

12. Repeat steps (1)-(11) until a convergence condition is fulfilled -- either a fixed number of iterations or stagnation of an error metric.

The processing time for 750 raw images with 1024 by 1024 pixels each is 2.5 mins for 3 iterations using a Dell XPS 8930 desktop computer. We validate the above recovery procedures via a simulation in Fig. 3. The black dots in Fig. 3(a) denote the initial guess of the positional shifts calculated via Eq. (3).

The red dots denote the ground-truth positions. Figure 3(b) shows the final recovered positional shift after the iterative refinement via Eq. (6). The mean error has been reduced from ~14 pixels to 0.19 pixels.

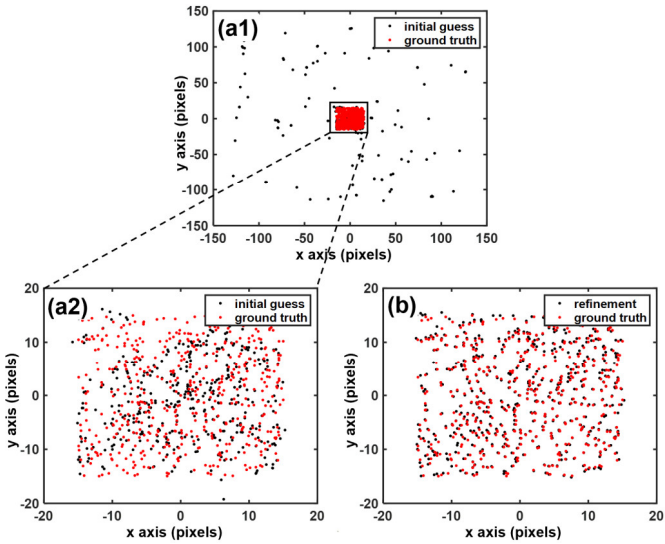

**Figure 3**. (a) The initial guess of the positional shifts (black dots). The mean error between the ground truth and the initial guess is ~14 pixels. (b) The recovered positional shifts after iterative refinement in the phase retrieval process. The mean error has been reduced to ~0.19 pixels.

## 4. Imaging demonstrations

In our experiments, we use a 532-nm fiber-coupled laser as the light source. We first validate the super-resolution imaging capability of the reported add-on module in Fig. 4. The sample is a USAF resolution target. The distance between the diffuser and the sample is ~0.5 mm. In our implementation, we first turn on the vibrational motors and then acquire the modulated raw images using a monochromatic camera (Sony IMX 250, 5-megapixel pixels, 75 fps). The acquisition time is set to be ~10 s, with a total number of ~750 images for reconstruction. We note that the image quality becomes better with an increased number of images, resulting in increased visibility of smaller features.

Figure 4(a) shows the incoherent summations of all acquired images and the diffraction-limited resolution is 4.38 μm half-pitch linewidth, corresponding to group 6, element 6 of the target. Figure 4(b) shows one captured raw image, where the speckle feature comes from the diffuser modulation process. Figure 4(c) shows the recovered super-resolution object and the diffuser profile. We can clearly resolve group 8, element 6 of the resolution target, corresponding to a 4-fold resolution gain over the diffraction limit of the employed objective lens. The current achievable resolution is limited by the feature size of the diffuser. One can use a monolayer of smaller particles to improve the modulation capability.

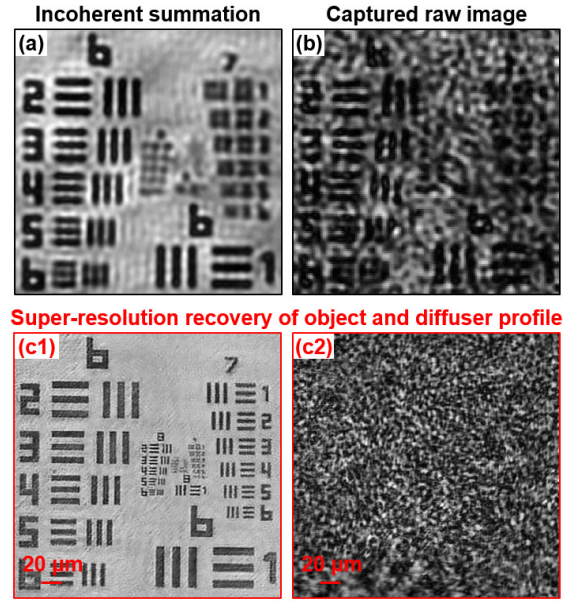

**Figure 4**. Test on an amplitude resolution target. (a) The incoherent summation of all captured images. (b) The captured raw image. (c) The recovered object and diffuser.

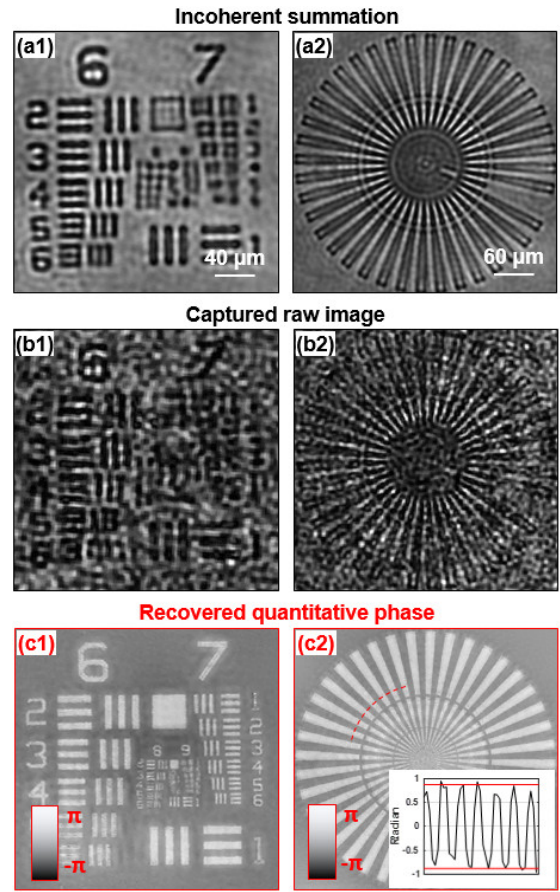

**Figure 5**. Test on a quantitative phase target. (a) The incoherent summation of all captured images. (b) The captured raw image. (c) The recovered quantitative phase.

In the second experiment, we test the PME add-on for quantitative phase imaging. A phase target (Benchmark QPT) is used as the object. Figure 5(a) shows the incoherent summation of all captured images. Figure 5(b) shows the captured raw image and Fig. 5(c) shows the recovered quantitative phase. The line profile across the red dash arc in Fig. 5(c2) is plotted in the inset, where the recovered phase is in a good agreement with the ground-truth phase.

for 54 and 78 μm. The distance of wavefront propagation is limited by the accuracy of the angular spectrum propagation method [49], which requires adequate sampling of the phase factor at the Fourier domain. For a thin sample, we can determine the best propagation distance by maximizing the image contrast or other metrics.

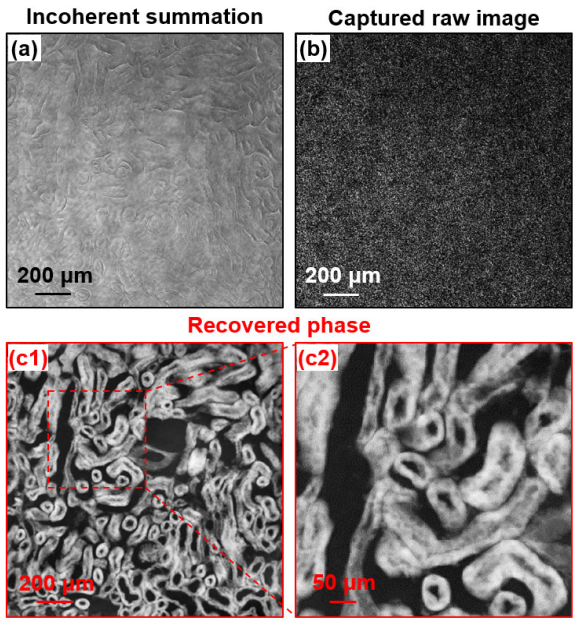

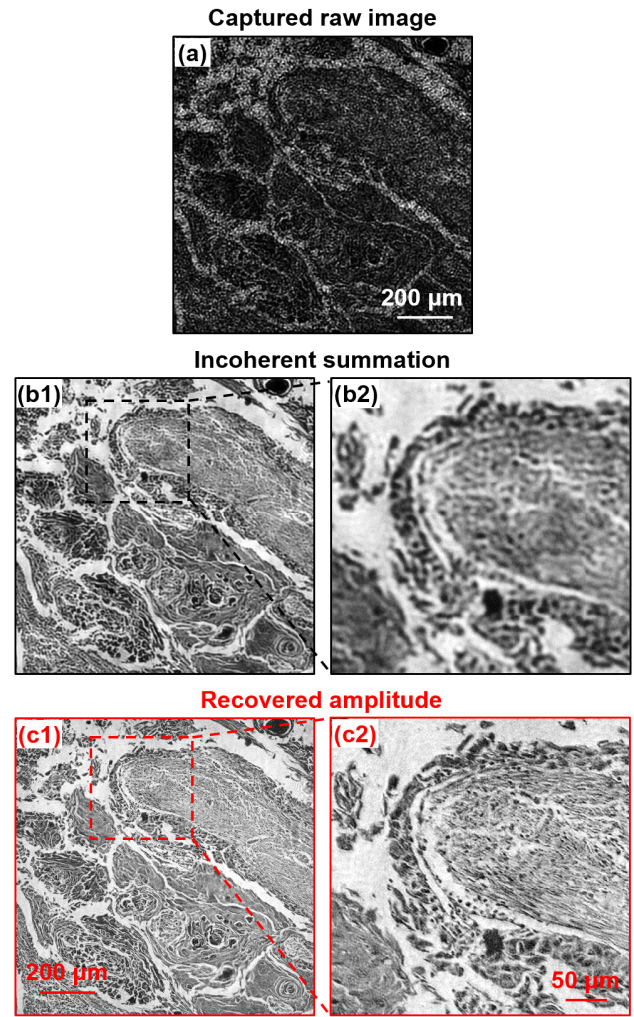

**Figure 6**. Test on an unstained mouse kidney slide. (a) The incoherent summation of all captured images. (b) The captured raw image. (c) The recovered phase.

In the third experiment, we test our add-on module on two fixed microscope slides: an unstained mouse kidney section (Fig. 6) and a Haemotoxylin and Eosin (H&E) stained breast tissue section (Fig. 7). Figure 6(a) shows the incoherent summation of all captured images of the unstained mouse kidney section. Figure 6(b) shows the captured raw image and Fig. 6(c) shows the recovered quantitative phase of the slide. Likewise, Fig. 7(a) shows the captured raw image of the H&E stained tissue section. Figure 7(b) shows the incoherent summation and Fig. 7(c) shows the super-resolution recovery.

In the fourth experiment, we test our add-on module for in-vitro yeast cell imaging. Figure 8(a) shows the recovered phase image of the yeast cells cultured on a solid agar plate. The incoherent summation of all captured images is shown in Fig. 8(b) for comparison. In this time-lapse experiment, we did not adjust the focus of the microscope stage. Figure 8(c1) and (d1) show the recovered phase images at time t = 3 h and 6 h. We can see that the recovered phase images are out of focus due to the axial drift of the microscope stage. One important advantage of the reported approach is to perform focusing after data has been acquired. Figure 8(c2) and (d2) show the refocused phase images after digitally propagating

**Figure 7**. Test on a stained breast tissue section. (a) The captured raw image. (b) The incoherent summation of all captured images. (c) The recovered amplitude.

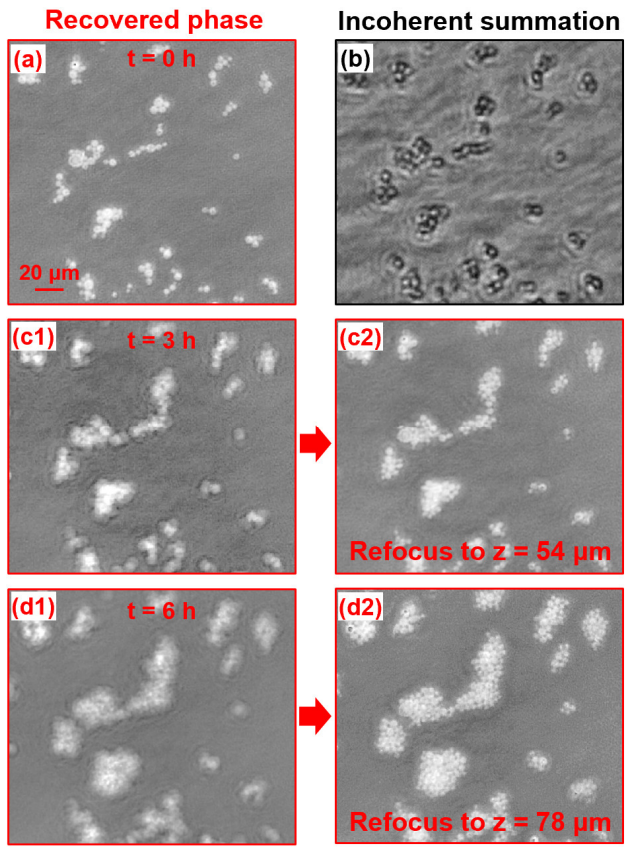

**Figure 8**. Test for in-vitro yeast cell imaging. (a) The recovered phase of the yeast cell at t = 0 h. (b) The incoherent summation of all captured images. (c1) The recovered phase image at t = 3 h. (c2) The refocused phase image at z = 54 µm. (d1) The recovered phase image at t = 6 h. (d2) The refocused phase image at z = 78 µm.

## 5. Discussion and conclusion

In summary, we report a simple and low-cost Ptychographic Modulation Engine (PME) add-on module for coherent super-resolution imaging. The reported add-on can be attached to most existing microscope platforms to perform large field-of-view and high-resolution ptychographic imaging. In our prototype setup, we demonstrate a 4-fold resolution over the diffraction limit of the employed 2X objective lens. We also test the module for in-vivo cell imaging where we can perform refocusing after the data has been acquired.

There are several unique advantages of the reported microscope add-on. First, we place the diffuser at a short distance to the sample. The speckle feature from the diffuser can be clearly resolved from the captured images. As such, we can directly recover the unknown positional shifts via image cross-correlation at each iteration step. In a conventional ptychography experiment, however, the image sensor is placed at the Fourier plane. It is challenging to recover the positional shift without a good initial guess. Second, the reported module is able to improve the resolution beyond the diffraction limit of the employed lens. This is similar to the Fourier ptychography approach using an LED array for angle-varied illumination. However, the Fourier ptychography relies on the thin sample requirement. The reported module, on the other hand, converts the thin-sample requirement to a thin-diffuser requirement. Once we recover the object wavefront, we can digitally propagate it back to any plane along the optical axis. The reported module may find applications in label-free bio-imaging where both large field-of-view and high resolution are needed. Third, the cost of the add-on module is no more than $5, providing a turnkey solution for most biologists and microscopists.

## Acknowledgments

Guoan Zheng acknowledges the support of National Science Foundation 1510077 and National Institute of Health R03EB022144.